\documentclass{aa}

\usepackage{graphicx}
\usepackage{amsmath}
\usepackage{amssymb}

\begin{document}
\title{Large amplitude oscillatory motion along a solar filament}

   \author{B.~Vr\v{s}nak\inst{1} \and
              A.~M.~Veronig\inst{2} \and
              J. K. Thalmann\inst{3} \and
              T. \v{Z}ic\inst{1}
           }
  \offprints{B.~Vr\v{s}nak, \email{bvrsnak@geof.hr}}
   \institute{
   Hvar Observatory, Faculty of Geodesy, Ka\v ci\'ceva 26, HR--10000 Zagreb, Croatia
   \and
   Institute of Physics, University of Graz, Universit\"atsplatz 5, A--8010 Graz, Austria
    \and
   Max-Planck-Institut f\"ur Sonnensystemforschung, Max-Planck-Str. 2, 37191 Katlenburg-Lindau, Germany
             }
   \date{Received 18 April 2007 / Accepted 22 May 2007}

 \abstract{}{}{}{}{}

  \abstract
     {Large amplitude oscillations of solar filaments is a phenomenon known for more than half a century.
     Recently, a new mode of oscillations, characterized by periodical plasma motions along
     the filament axis, was discovered. }
     {We analyze such an event, recorded on 23 January 2002 in Big Bear Solar Observatory
     H$\alpha$ filtergrams, in order to infer the triggering mechanism
     and the nature of the restoring force.}
     {Motion along the filament axis of a distinct buldge-like feature was traced, to
     quantify the kinematics of the oscillatory motion. The data were fitted
     by a damped sine function, to estimate the basic parameters of the oscillations.
     In order to identify the triggering mechanism, morphological changes in the vicinity
     of the filament were analyzed.}
     {The observed oscillations of the plasma along the filament was characterized by
     an initial displacement of 24~Mm, initial velocity amplitude of 51~km\,s$^{-1}$,
      period of 50~min, and damping time of 115 min.
     We interpret the trigger in terms of poloidal magnetic flux injection by magnetic reconnection at one of the
     filament legs. The restoring force is caused by the magnetic pressure gradient along the
     filament axis. The period of oscillations, derived from the linearized equation of motion
     (harmonic oscillator) can be expressed as
     $P=\pi\sqrt{2}L/v_{A\varphi}\approx4.4L/v_{A\varphi}$, where
     $v_{A\varphi} =B_{\varphi0}/\sqrt{\mu_0\rho}$ represents the Alfv\'en speed based on
     the equilibrium poloidal field $B_{\varphi0}$.}
     {Combination of our measurements with some previous observations of the same kind
     of oscillations shows a good agreement with the proposed interpretation.}


   \maketitle

\section{Introduction}

Solar coronal structures are frequently subject to oscillations of various modes and time/spatial scales
(e.g., Aschwanden~\cite{aschw03}). The longest known phenomenon of this kind are so called winking filaments
(cf., Ramsey \& Smith~\cite{R&S66}). They represent large-amplitude large-scale oscillations of prominences
observed on the solar disc, most often triggered by disturbances coming from distant flares. Later on,
various modalities of prominence oscillatory motions were reported (e.g., Kleczek \& Kuperus~\cite{K&K69},
Malville \& Schindler~\cite{M&S81}, Vr\v{s}nak~\cite{vrs84}, Wiehr et al.~\cite{wiehr84}, Tsubaki \&
Takenchi~\cite{T&T86}, Vr\v{s}nak et al.~\cite{vrs90}, Jing et al.~\cite{jing03}, Isobe \&
Tripathi~\cite{I&T06}). The phenomenon is generally interpreted in terms of different magnetohydrodynamical
(MHD) wave modes (for a classification we refer to reviews by Tsubaki~\cite{tsubaki88},
Vr\v{s}nak~\cite{vrs93}, Roberts~\cite{rob00}, Oliver \& Ballester~\cite{O&B02}).

Most of reported large-amplitude large-scale oscillations happen perpendicular to the prominence axis. The
restoring force in this type of oscillations generally could be explained in terms of the magnetic tension
(e.g., Hyder~\cite{hyder66}, Kleczek \& Kuperus~\cite{K&K69}, Vr\v{s}nak~\cite{vrs84}, Vr\v{s}nak et
al.~\cite{vrs90}). On the other hand, the oscillation damping in the corona is attributed either to the
``aerodynamic" drag (i.e., the energy loss by emission of waves into the ambient corona; e.g., Kleczek \&
Kuperus~\cite{K&K69}), or to various dissipative processes  (e.g., Hyder~\cite{hyder66}, Nakariakov et
al.~\cite{nakar99}, Terradas et al.~\cite{terradas01}, Ofman \& Aschwanden~\cite{O&A02}, Verwichte et
al.~\cite{verwichte04}).

Recently, Jing et al.~(\cite{jing03,jing06}) reported a new mode of oscillations in filaments/prominences,
where the oscillatory motion happens along the prominence axis. In this paper we report another example
observed in H$\alpha$ filtergrams and propose an explanation for the triggering process and the restoring
force.

\section{Observations }

\begin{figure*}
\centering \resizebox{0.75\hsize}{!}{\includegraphics{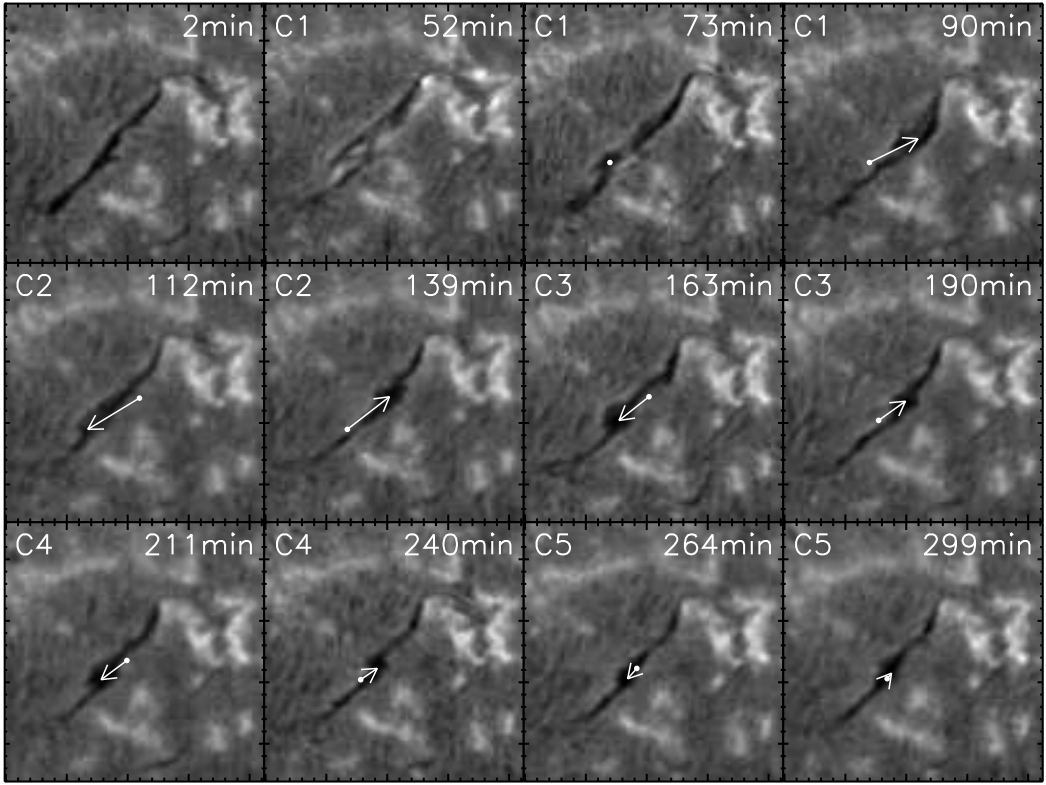}}
 \caption{Sequence of BBSO H$\alpha$ images for a $210'' \times 210''$ subfield.
 For each image, the observing time (in minutes after 17:00~UT) is annotated.
 The arrows connect the determined displacement for consecutive images
taken around the oscillation peaks for each of the five cycles (C1-C5).
  The arrowhead indicates the determined center of the followed filament
feature in the actual image, the circle drawn at the other end of the arrow indicates the location of the
same feature as determined from the previous image shown in the figure. Thus, the decreasing lengths of the
arrows directly reflect the instantaneous oscillation amplitudes and their damping.
 See also the accompanying movie.
     \label{ha_series} }
\end{figure*}

The filament that we discuss was located near the center of the solar disk on 2002 January 23 and was
observed in full-disk H$\alpha$ filtergrams at 1$''$/pixel resolution with a time cadence of 1~min at the Big
Bear Solar Observatory (BBSO) Global H$\alpha$ Network Station (Steinegger et al.~\cite{stein00}). The
accompanying movie shows the evolution of the oscillating filament between 17 and 22 UT; snapshots are
presented in Fig.~\ref{ha_series}. During this time span, the filament was activated and showed oscillations
along its main axis which could be followed for 5~cycles.

Before the measurements, we co-aligned all the images to the reference image taken at 17:00~UT (hereinafter
used as $t=0$), correcting for solar differential rotation and also for image jittering by the means of
cross-correlation techniques. For measuring the oscillatory motions of the filament, we did not use all the
available images but on average each third image of the series, giving an average time cadence of 3~min.

\begin{figure*}
\centering \resizebox{0.98\hsize}{!}{\includegraphics{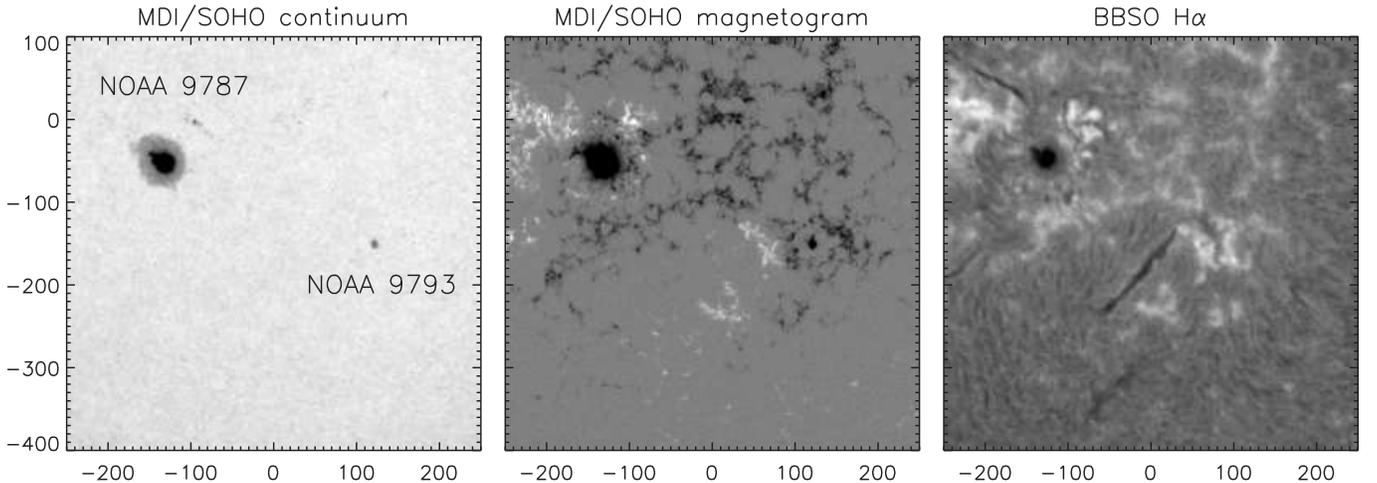}}
 \caption{MDI/SOHO continuum image, MDI/SOHO magnetogram and BBSO H$\alpha$ filtergram at 17~UT
 of active regions NOAA 9787 and NOAA 9793 with the filament. Units of the x- and y-axis are in arc sec.
    \label{ha_mdi} }
\end{figure*}

The main part of the filament, 110\,--\,120 Mm long, was following the straight magnetic inversion
line oriented in Southeast-Northwest direction. Its Northwest leg, a curved thread of length
20\,--\,30~Mm, was rooted in a small active region, NOAA 9793 (Fig.~\ref{ha_mdi}). The Southeast
leg was rooted in the quiet chromosphere. For the total length of the filament we take $2L=140$~Mm.

\begin{figure}
\centering
 \resizebox{0.9\hsize}{!}{\includegraphics{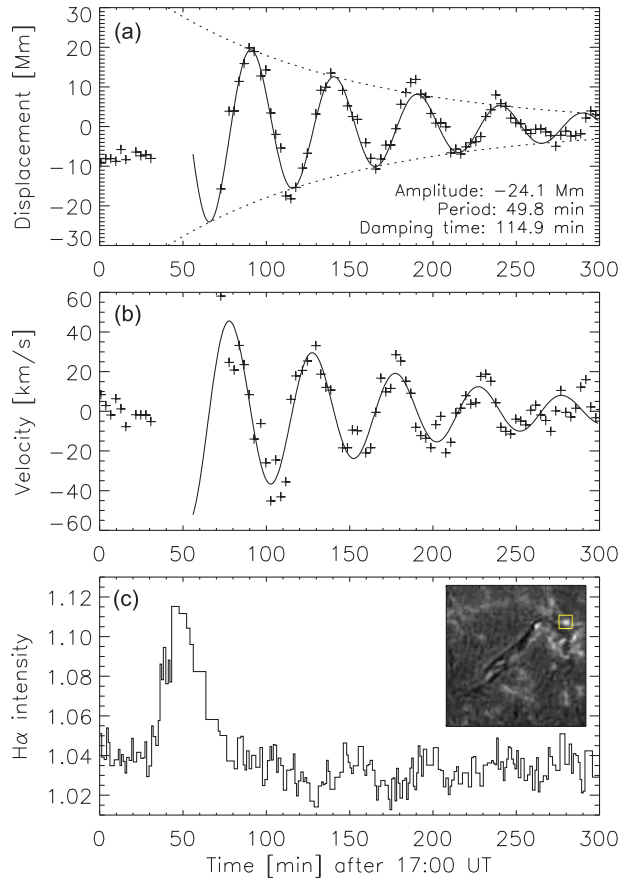}}
 \caption{a) Displacement measurements of the oscillating filament together
 with the applied fit. b) Velocities derived
 via numerical differentiation from the displacement measurements (for the procedure see main text)
 shown together with the time-derivative of the $d(t)$ fit shown in panel a). c) H$\alpha$ light curve measured in the
 western footpoint of the associated subflare indicated by the box in the inserted image.
     \label{osc_plot1} }
\end{figure}

The kinematics of the observed oscillations was determined by visually inspecting the H$\alpha$
image sequence, and measuring the displacement along the filament axis of the most prominent (dark)
bulge-like feature in the filament (Fig.~\ref{ha_series}; see also the accompanying movie). The
measured displacements are presented in Fig.~\ref{osc_plot1}a, and the corresponding velocities in
Fig.~\ref{osc_plot1}b.
 Note that plasma motions could be seen all along the filament, with the amplitude decreasing
towards the filament legs, i.e., the oscillation amplitude was largest in the middle part of the filament,
where the bulge-like feature was located. A mechanical analogue of the observed oscillation would be, e.g., a
longitudinal-mode standing wave on a slinky spring fixed at both ends.

It seems that the filament oscillation was a response to an energy release in the active region, where a weak
flare-like brightening appeared in H$\alpha$ at 17:33~UT in the region of the Northwest footpoint of the
filament (see the second panel in the first row of Fig.~\ref{ha_series}; see also the accompanying movie).
The H$\alpha$ subflare achieved a maximum around 17:45~UT, as can be seen in Fig.~\ref{osc_plot1}c where we
show the H$\alpha$ light curve, measured at the western kernel of the subflare (see the inset in
Fig.~\ref{osc_plot1}c; the eastern kernel was partly obscured by the activated filament, thus it was not
possible to measure the lightcurve in this part of the subflare. The energy release associated with the
H$\alpha$ brightening was obviously weak, since no increase in the GOES full-disk integrated soft X-ray flux
was observed.

The subflare caused the filament activation, where various forms of motions and restructuring could
be observed. Due to complex motions and morphological changes of the filament, in the period
17:35\,--\,18:10 it was practically impossible to identify, and measure reliably, the position of
the filament feature which was later on used to trace the oscillations. In the period between 18:00
and 18:10~UT the motions gradually became more ordered, mainly showing a flow of the filament
plasma in the Southeast direction, away from the active region. After this initial plasma
displacement, the first Northwest directed swing started around 18:10~UT (Fig.~\ref{osc_plot1}a;
the first oscillation measurement was performed at 08:12~UT).

The measured displacement curve, $d(t)$, of the oscillatory filament motion was fitted by a damped sine
function of the form
\begin{equation}
\tilde{d}(t) = x_0 \sin \left( \frac{2\pi t}{P} + \varphi \right)  \exp\left( -\frac{t}{\tau} \right) + d_0
\end{equation}
with amplitude~$x_0$, period $P$, phase $\varphi$, damping time~$\tau$ and equilibrium position
$d_0$. Fig.~\ref{osc_plot1}a shows the observed oscillatory motion of the filament (with $d_0$
already subtracted), together with the fitted curve. In Fig.~\ref{osc_plot1}b we plot the
corresponding velocities obtained by numerical differentiation of the displacement data points
(applying a 3-point, Lagrangian interpolation) together with the time derivative of the $d(t)$ fit.

The basic parameters of the oscillations are:
 \begin{itemize}
\item period $P=49.8$~min;
\item initial amplitude $x_0=24.1$~Mm;
\item initial velocity amplitude $v_0=51$~km\,s$^{-1}$;
\item damping time $\tau=114.9$~min.
 \end{itemize}

\section{Interpretation and discussion}

The oscillations described by Eq.~(1) are the solution of the equation of motion of the form:
 \begin{equation}
 \ddot x + 2\delta\dot x + \omega_0^2 x = 0\,.
 \end{equation}
The cyclic frequency of oscillations $\omega$ is related to the frequency of free oscillations $\omega_0$ and
the damping rate $\delta$ by the expression $\omega^2=\omega_0^2-\delta^2$. From the period of oscillations
and the damping time determined in Sect.~2 one finds the cyclic frequency of oscillations and the damping
rate as $\omega=2\pi/P=2.1\times10^{-3}$~s$^{-1}$ and $\delta=1/\tau=1.5\times10^{-4}$~s$^{-1}$,
respectively. Since $\omega\gg\delta$, it can be concluded that $\omega_0\approx\omega$, so in the following
we take for the eigenmode frequency of the system the value $\omega_0\approx2\times10^{-3}$~s$^{-1}$. From
the cyclic frequency $\omega_0$ and the amplitude $x_0$ one finds the initial velocity and acceleration
amplitudes, $v_0=\omega_0 x_0= 50.8$~km\,s$^{-1}$ and $a_0=\omega_0^2x_0=107$~m\,s$^{-2}$.

The estimated value of acceleration gives a constraint to the nature of the restoring force. Let us first
assume that the restoring force is governed by the gas pressure gradient. It can be approximately estimated
as $\Delta p/2L$, where we assume that there is a pressure difference $\Delta p$ over the filament length
$2L$. Such a pressure gradient would give rise to an acceleration of $a=\Delta p/2L\rho$, where $\rho$ is the
prominence density. Writing down an order of magnitude estimate $2La=\Delta p/\rho\approx p/\rho\approx
c_s^2$, one finds that the sound speed has to be $c_s=\sqrt{2La}\gtrsim100$~km\,s$^{-1}$. Such a sound speed
requires temperatures $\gtrsim10^6$~K, which is typical for the coronal plasma. So, the required sound speed
is an order of magnitude too high, since the temperature of the prominence plasma is $\approx$\,$10^4$~K,
corresponding to a sound speed in the order of $c_s\approx10$~km\,s$^{-1}$. A similar conclusion could be
drawn by taking into account that the period of oscillation can be roughly identified with the sound wave
travel time along the filament axis, $t_s\approx 2L/c_s$, i.e., expressing the period as $P\approx
2t_s=4L/c_s$. Substituting $2L=140$~Mm, we find again $c_s>100$~km\,s$^{-1}$
(note that using the oscillation amplitude  $x_0$ instead of the filament length 2$L$
 would give an estimate of the
 velocity amplitude and not the oscillation-related wave velocity, i.e., the sound speed).

So, the pressure pulse scenario would be possible only if the pressure of the filament plasma was increased
by a factor of 100. That corresponds to a temperature increase from $T_0\approx 10^4$~K to $T\gtrsim10^6$~K,
but no signature of such plasma heating was observed in the TRACE 171~{\AA} ($T \sim 1$~MK) EUV images.
Another possibility is that the restoring force is of magnetic origin. For example, let us consider that the
filament is embedded in a flux rope and that at a certain moment an additional poloidal flux is injected into
the rope at one of its legs. The flux could be injected from below the photosphere, or by magnetic field
reconnection (see, e.g., Uchida et al.~\cite{uchida01}, Jibben \& Canfield~\cite{J&C04}). Indeed, H$\alpha$
flare-like brightenings appeared just before/around the onset of oscillations in the region of the Northwest
footpoint of the filament (see Sect.~2), which is likely to be a signature of reconnection. On the other
hand, no sign of the emerging flux process could be detected in the region around the filament in the
SOHO/MDI magnetograms which were available at a time cadence of 1~min.

The excess poloidal field creates a magnetic pressure gradient and a torque, driving a combined
axial and rotational motion of plasma (e.g., Jockers~\cite{jock78}, Uchida et al.~\cite{uchida01}).
If the injection lasts shorter than the Alfv\'en travel time along the flux rope axis, it causes a
perturbation that propagates along the rope (Uchida et al.~\cite{uchida01}) and evolves into an
oscillatory motion along its axis. Here we assume that the wave caused by the injection is
reflected at the opposite footpoint of the rope, and that the damping rate is smaller than the
oscillation cyclic frequency. Also note that since $c_s<v_A$ the plasma is carried by $B_{\varphi}$
since plasma motions along the field lines are slower than translation of poloidal field (fast-mode
MHD wave).

Let us consider a flux rope of a constant width, where a certain amount of poloidal flux is
injected at one of its footpoints (similar to that shown in Fig.~5 of Jibben \&
Canfield~\cite{J&C04}), over an interval shorter than the eigenmode period of the related
oscillation mode. We divide the flux rope in equilibrium into two segments of length $L$, symmetric
with respect to the flux rope midpoint. In the simplest form, the average poloidal magnetic field
in the two segments, after perturbing the flux rope, can be expressed as
$B_{\varphi1}=B_{\varphi0}L/(L-x)$ and $B_{\varphi2}=B_{\varphi0}L/(L+x)$. Here, $B_{\varphi0}$ is
the poloidal field in the equilibrium state and $x$ is the longitudinal displacement of the plasma
element separating the two segments (in the equilibrium taken to be at the filament barycenter,
located at the midpoint of the flux rope). We use these relations to express the corresponding
magnetic pressures, $p_m=B_{\varphi}^2/2\mu_0$, which we substitute into the simplified equation of
motion:
 \begin{equation}
 \rho\ddot x  \approx -\frac{\Delta p_m}{L}=
   - \frac{B^2_{\varphi0}}{2\mu_0} \left[\frac{1}{(L-x)^2} - \frac{1}{(L+x)^2}\right]L\,,
 \end{equation}
where $\Delta p_m/L$ represents approximately the gradient of the poloidal-field magnetic pressure. After
linearizing Eq.~(3), and expressing the displacement in the dimensionless form $X=x/L$, we get the equation
of motion of the harmonic oscillator:
 \begin{equation}
 \ddot X  = - \frac{2v_{A\varphi}^2}{L^2}~  X\,,
 \end{equation}
where $v_{A\varphi} =B_{\varphi0}/\sqrt{\mu_0\rho}$ represents the Alfv\'en speed based on the equilibrium
poloidal field $B_{\varphi0}$.

Equation~(4) implies $\omega_0^2=2\,v_{A\varphi}^2/L^2$, and consequently,
$P=\pi\sqrt{2}L/v_{A\varphi}\approx4.4L/v_{A\varphi}$. Utilizing the observed parameters, we find
$v_{A\varphi}\approx 100$~km\,s$^{-1}$. Assuming that the number density of the prominence plasma
is in the range $n=10^{10}$\,--\,$10^{11}$\,~cm$^{-3}$ (cf. Tandberg-Hanssen~\cite{tandb95}), one
finds $B_{\varphi0}\approx 5$\,--\,15 gauss.

In this respect we note the filament was composed of helical-like fine structure patterns, twisted around the filament
axis. Bearing in mind that the frozen-in condition is satisfied in the prominence plasma, such patterns are
indicative of the flux-rope helical magnetic field (Vr\v{s}nak et al.~\cite{vrs91}). In the period
18:25\,--\,18:40~UT the internal structure was seen clearly enough to provide measurements of the pitch angle
of helical threads. The measured values of the pitch angle range between 20$^{\circ}$ and 30$^{\circ}$, with
the mean value $\tan\vartheta=X=24^{\circ}\pm4^{\circ}$. Bearing in mind $X=B_{\varphi}/B_{\parallel}$, where
$B_{\varphi}$ and $B_{\parallel}$ are the poloidal and axial magnetic field components (Vr\v{s}nak et
al.~\cite{vrs91}), and considering the poloidal field of 5\,--\,15 gauss, one finds
$B_{\parallel}\approx10$\,--\,30~gauss, which is a reasonable value for quiescent prominences
(Tandberg-Hanssen~\cite{tandb95}).

According to Vr\v{s}nak~(\cite{vrs90}), who treated the stability of a semitoroidal flux
rope anchored at both legs in the photosphere, the prominence with pitch angle
$\vartheta=20^{\circ}$\,--\,$30^{\circ}$ should be stable, which is consistent with the fact that
the filament was severely activated, but did not erupt. A similar conclusion can be drawn if the
total number of turns of the helical field line is considered. Taking into account that the
filament width was around $2r\approx 2$~arcsec and employing the expression for the pitch-length
$\lambda=2\pi r/X$, we find the total number of turns of the helical field line
$N=2L/\lambda\approx 1.4$. This is again consistent with the model stability-criteria by
Vr\v{s}nak~\cite{vrs90} (for the unstable cases of highly twisted quiescent prominences see, e.g.,
Vr\v{s}nak et al.~\cite{vrs91} or Karlick\'y \& \v{S}imberov\'a~\cite{K&S02}). However, it should
be noted that sometimes active region prominences erupt at smaller number of turns ($N\gtrsim1$;
see, e.g., Rust \& Kumar~\cite{R&K96}, R\'egnier \& Amari~\cite{R&A04}, Salman et
al.~\cite{salman07}).


Regarding the $P\propto L$ relationship, we have to discuss Fig.~5a of Jing et al.~(\cite{jing06}), where in
the figure caption it is stated that the $x$-axis of the graph represents the ``length". However, checking
Table I therein, one finds that in fact the $x$-axis represents the oscillation amplitude, which we
previously denoted as $x_0$. Consequently, the lack of correlation does not really mean that there is no
$P\propto L$ relationship. In this respect it is instructive to comment also Fig.~5b of Jing et
al.~(\cite{jing06}) which shows proportionality of the velocity amplitude $v_0$ and ``length", the latter in
fact representing the amplitude $x_0$. Such a proportionality is expected for any harmonic oscillator since
$v_0=\omega_0 x_0$. Bearing in mind Eq.~(4), i.e., $\omega_0\propto 1/L$, one would expect an inverse
proportionality between $v_0$ and $L$ in the case of equal amplitudes.

Out of four events studied by Jing et al.~(\cite{jing03,jing06}), in two cases it is possible to estimate the
overall length of the oscillating filaments. From filtergrams displayed in Fig.~1 of Jing et
al.~(\cite{jing03}) and Fig.~3 of Jing et al.~(\cite{jing06}), showing filaments characterized by periods
$P=80$ and 100~min, we estimated the filament lengths to approximately $2L\approx200$~Mm and
$2L\approx300$~Mm, respectively. Combining these values with our measurements ($P=50$~min;
$2L\approx140$~Mm), we find that the data follow approximately the $P\propto L$ dependence. The linear least
squares fit gives a slope of 0.6~min/Mm, or 0.7~min/Mm if the fit is fixed at the origin. If we compare these
values with the previously derived relationship $P\approx4.4L/v_{A\varphi}$, we find that these slopes
correspond to $v_{A\varphi}\approx 100$\,--\,120~km\,s$^{-1}$, which is in agreement with the value of
$v_{A\varphi}$ that we estimated earlier for our oscillating filament.

Finally, let us note that the poloidal flux injection changes the poloidal-to-axial field ratio
$X=B_{\varphi}/B_{\parallel}$, which should cause also oscillations perpendicular to the prominence axis,
since the equilibrium height of the prominence is directly related to the ratio $X$ (e.g.,
Vr\v{s}nak~\cite{vrs84,vrs}).
Indeed, careful inspection of the filament motion reveals such transversal oscillations of a small
amplitude, synchronous with the longitudinal ones (see the accompanying movie). The analysis of the
interplay between the longitudinal, transversal, torsional (pitch-angle), and radial oscillations
requires a meticulous treatment, which will be presented in a separate paper.

\section{Conclusion}

The observed oscillations of the plasma along the filament was characterized by a period of 50~min,
velocity amplitude of 50~km\,s$^{-1}$, and damping time of 115~min. Our analysis indicates that the
oscillations of the filament plasma along its axis were driven magnetically as in other types of
prominence oscillations. We propose that the oscillations were triggered by an injection of
poloidal magnetic field into the flux rope, most probably by the reconnection associated with a
subflare that took place at the Northwest leg of the filament. The flux injection is expected to
cause a translatory and torsional motion of plasma, propagating toward the other leg of the
filament. After the reflection of the wave, the oscillations along the filament axis develop,
showing an order of magnitude weaker damping than in perpendicular oscillations of filaments (for
the latter see, e.g., Hyder~\cite{hyder66}).
Most probably, perpendicular oscillations decay much faster because they
affect the ambient corona more than motions along the flux tube, i.e., the energy flux carried
away by MHD waves is much larger.

The period in our event is shorter than in analogous oscillations analyzed by Jing et
al.~(\cite{jing03,jing06}), where the periods were in the range from 80 to 160~min. In two of these events
the filament length could be measured, and when supplemented with our event, we find that the periods follow
the $P\propto L$ relationship. The slope of the relationship is consistent with a poloidal magnetic field in
the order of 10~gauss.

The damping time $\tau$ is also shorter than in the events analyzed by Jing et
al.~(\cite{jing03,jing06}). If our measurements would be inserted into Fig.~5c of Jing et
al.~(\cite{jing06}), the corresponding data-point would be located very close the lower end of the
regression line presented therein. The linear least squares fit, with the axis intercept fixed at
the origin, would read $\tau=3.5P$.

\section*{Appendix: The nonlinearity aspect}

\begin{figure}
\centering
 \resizebox{0.9\hsize}{!}{\includegraphics{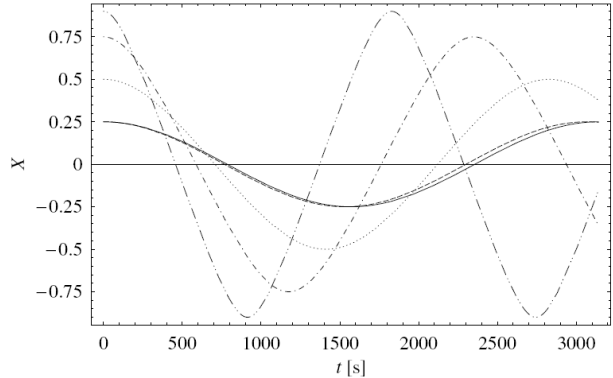}}
 \caption{Solutions of Eq.~(5) shown for amplitudes $X_0=0.25$, 0.5, 0.75, and 0.9
         (dashed, dotted, dot-dashed and dot-dot-dashed line respectively),
           together with the harmonic oscillator solution for $X_0=0.25$ (full line).
     \label{nonlin} }
\end{figure}

Since the filament oscillations were characterized by large amplitudes, it is instructive to check
how much the harmonic-oscillator approximation differs from the solutions of Eq.~(3). After
introducing the dimensionless displacement $X=x/L$, Eq.~(3) can be rewritten in the form:
 \begin{equation}
 \ddot X  =
   - \frac{\omega_0^2 X}{(1-X^2)^2} \,,
 \end{equation}
where $\omega_0^2=2\,v_{A\varphi}^2/L^2$. In Fig.~\ref{nonlin} we show the solutions of Eq.~(5) for
amplitudes $X_0=0.25$, 0.5, 0.75, and 0.9 together with the harmonic oscillator solution for $X_0=0.25$. We
have chosen $\omega_0^2=4\times10^{-6}$~s$^{-1}$, approximately corresponding to $P=50$~min. Inspecting
Fig.~\ref{nonlin} one finds that the period decreases with increasing amplitude: it is about 2\,\% shorter
than the harmonic oscillator period for $X_0=0.25$, around 10\,\% shorter at $X_0=0.5$, and 25\,\% at
$X_0=0.75$. The deviation from the harmonic oscillator period at $X_0=0.9$ becomes larger than 40\,\%.
Bearing in mind the damping, the increase of the period with decreasing amplitude means that the period
should increase as the oscillations attenuate. Indeed, going back to Fig.~\ref{osc_plot1}, we find that the
measurements might indicate such a trend towards the end of oscillations, in particular after $t>250$~min.

\acknowledgements

We are thankful to Vasyl Yurchyshyn for providing us with the data from the Global High Resolution H$\alpha$
Network, operated by the Big Bear Solar Observatory, New Jersey Institute of Technology. We are grateful to
the referee, Dr. Marian Karlick\'y, whose constructive suggestions led to significant improvements in this
paper. Travel grants from the exchange program Austria-Croatia WTZ 08/2006 are gratefully acknowledged.

\end{document}